\documentclass[runningheads]{llncs}
\usepackage[T1]{fontenc}

\usepackage{graphicx}

\usepackage{caption}
\usepackage{multirow}
\usepackage[table,xcdraw]{xcolor}
\usepackage{algorithm}
\usepackage{booktabs}
\usepackage{enumitem}
\usepackage{algpseudocode}
\floatname{algorithm}{Algorithm} %
\usepackage{amsmath,amsfonts,amssymb}
\usepackage{float} 
\usepackage{adjustbox}
\usepackage{subfig}

\usepackage{makecell}

\usepackage{siunitx}
\usepackage{tabularx}

\usepackage[misc]{ifsym} %
\usepackage{color}
\usepackage[
  colorlinks=true,
  linkcolor=blue,   %
  citecolor=blue   
]{hyperref}

\begin{document}
\title{DynamicPO: Dynamic Preference Optimization for Recommendation}

%
%

\author{Xingyu Hu\inst{1,2}\textsuperscript{$\dagger$(\Letter)} \and
Kai Zhang\inst{1} \and
Jiancan Wu\inst{1}\textsuperscript{(\Letter)} \and
Shuli Wang\inst{3} \and
Chi Wang\inst{3} \and
Wenshuai Chen\inst{3} \and
Yinhua Zhu\inst{3} \and
Haitao Wang\inst{3} \and
Xingxing Wang\inst{3} \and
\mbox{Xiang Wang}\inst{1}}

\authorrunning{X. Hu et al.}

\institute{University of Science and Technology of China, Hefei, China \\
\email{huxy@mail.ustc.edu.cn, \{kaizhang1215, wujcan, xiangwang1223\}@gmail.com} 
\and
Shanghai Innovation Institute, Shanghai, China
\and
Meituan, Chengdu, China \\
\email{\{wangshuli03, wangchi06, chenwenshuai, zhuyinhua, wanghaitao13, wangxingxing04\}@meituan.com} }

\maketitle              
\begingroup
\let\thefootnote\relax
\footnotetext{\textsuperscript{$\dagger$}Work done at Meituan.}
\endgroup

\begin{abstract}
In large language model (LLM)-based recommendation systems, direct preference optimization (DPO) effectively aligns recommendations with user preferences, necessitating multi-negative objective functions to leverage abundant implicit-feedback negatives and to sharpen preference boundaries. However, our empirical analyses reveal a counterintuitive phenomenon—\emph{preference optimization collapse}—where increasing the number of negative samples can lead to performance degradation despite a continuously decreasing training loss. We further theoretically demonstrate that this collapse arises from \emph{gradient suppression}, caused by the dominance of easily discriminable negatives over boundary-critical negatives that truly define user preference boundaries, leaving boundary-relevant signals under-optimized and ultimately weakening the model's decision boundary. Motivated by these observations, we propose \textbf{DynamicPO} (\textbf{Dynamic} \textbf{P}reference \textbf{O}ptimization), a lightweight and plug-and-play framework comprising two adaptive mechanisms: (1) \emph{Dynamic Boundary Negative Selection}, which identifies and prioritizes informative negatives near the model's decision boundary, and (2) \emph{Dual-Margin Dynamic $\beta$ Adjustment}, which calibrates optimization strength per sample according to boundary ambiguity. Extensive experiments on three public datasets show that DynamicPO effectively prevents optimization collapse and improves recommendation accuracy on multi-negative preference optimization methods, with negligible computational overhead.  Our code and datasets are available at \url{https://github.com/xingyuHuxingyu/DynamicPO}.
\end{abstract}


\section{Introduction}

Sequential recommendation systems aim to predict the next item a user is likely to interact with by modeling their historical behaviors \cite{sequential_survey}. With the advent of large language models (LLMs), which possess extensive world knowledge and powerful reasoning capabilities, the landscape of recommender systems is rapidly evolving. Recent studies have shown that LLMs can serve as a strong backbone for next-generation recommendation algorithms, outperforming traditional architectures in various domains \cite{sequential_rec_llm,wu2024llm_rec_survey,bao2023tallrec,liao2024llara}.

Despite recent progress in leveraging large language models (LLMs) for recommendation tasks, the majority of current approaches predominantly employ supervised fine-tuning (SFT) based on language modeling objectives. Specifically, these methods train models to predict the next token or to continue sequences of text, which primarily encourages the generation of coherent and contextually appropriate language. However, this training paradigm does not explicitly guide the model to discern or rank user preferences, as it focuses on sequence continuation rather than on preference discrimination. Consequently, such objectives are inherently misaligned with the goal of recommendation, which is to accurately rank items according to users' likes and dislikes. This misalignment may limit the effectiveness of LLM-based recommendation models in capturing nuanced user preferences and delivering truly personalized recommendations \cite{DMPO}.

\begin{figure}[htbp]
  \centering
  \subfloat[Preference optimization of LLM-based Recommenders.\label{fig:teaser_collapse_phenomenon}]{
    \includegraphics[width=0.48\textwidth]{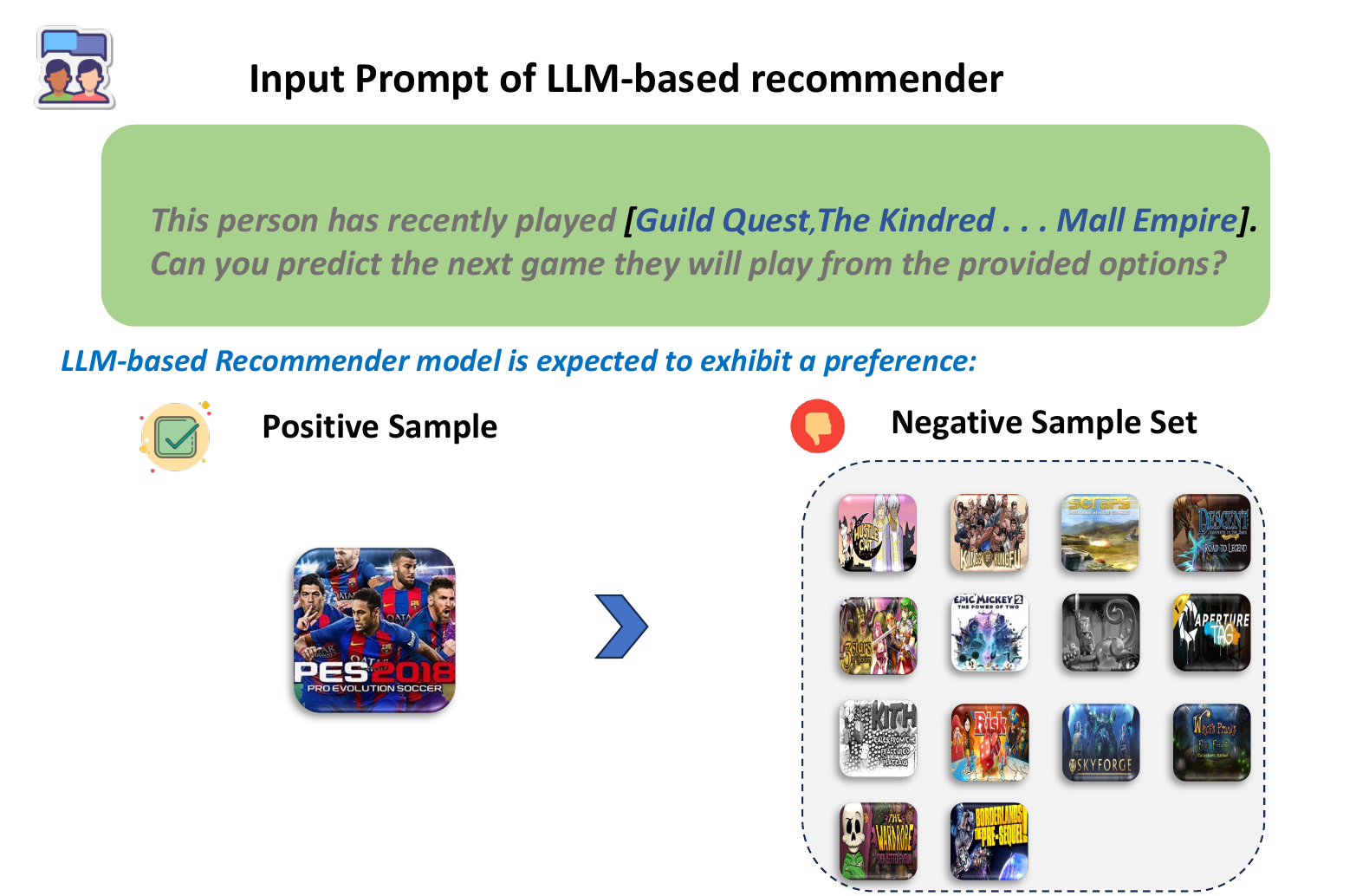}
  }\hfill
  \subfloat[The preference optimization collapse phenomenon observed in DMPO.\label{fig:dmpo_collapse}]{
    \includegraphics[width=0.48\textwidth]{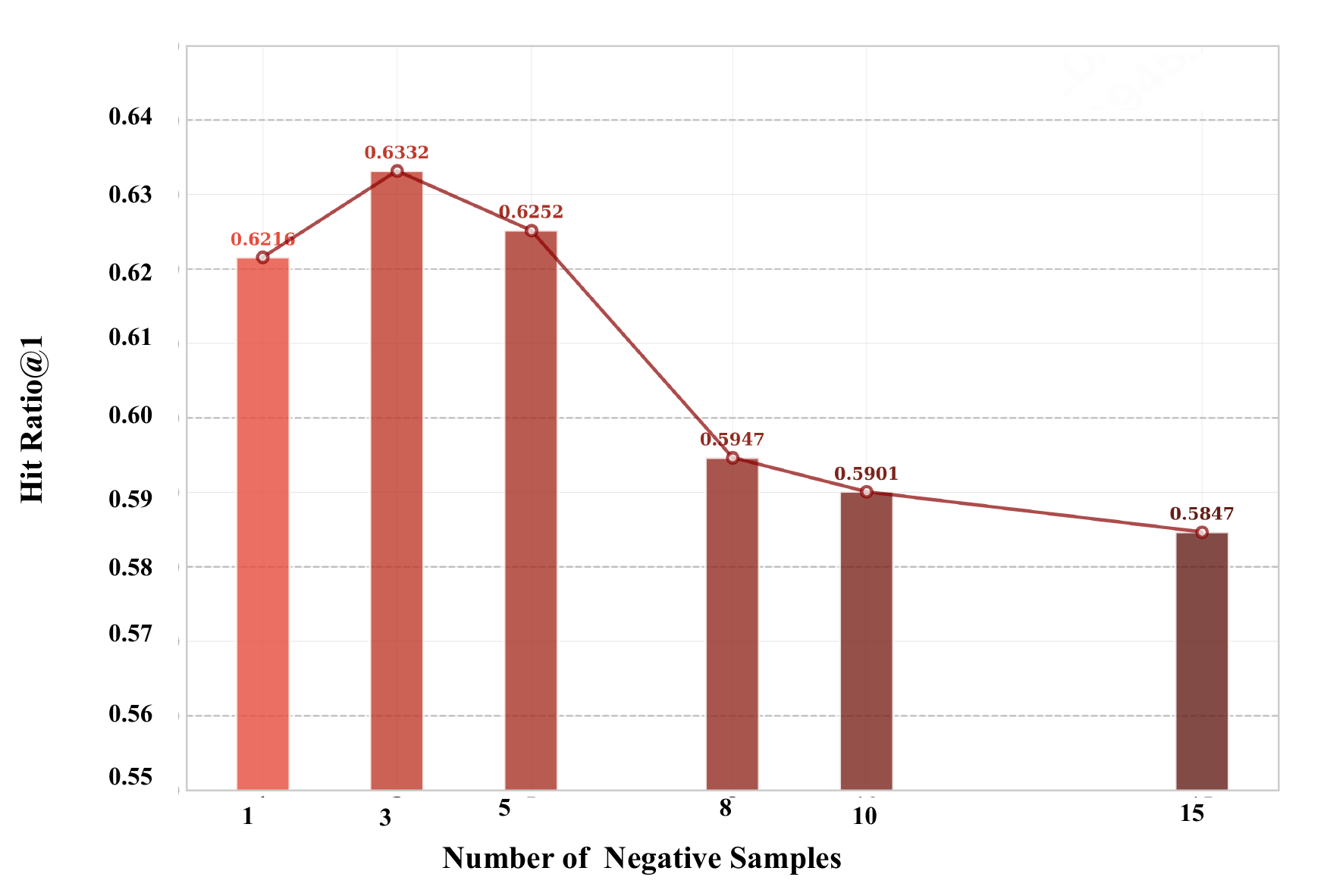}
  }
  \caption{Demonstration of preference optimization and collapse phenomenon in LLM-based recommenders.}
\end{figure}

To bridge this gap, preference optimization techniques—originally developed for dialogue and instruction tuning—have been adapted to the recommendation domain. These methods, especially in two-stage pipelines (first distilling item-level knowledge via SFT, then refining subtle user preferences through preference optimization), have demonstrated substantial and consistent improvements over SFT alone \cite{DMPO}. The process trains models to prioritize the positive sample while deprioritizing negatives (see Figure~\ref{fig:teaser_collapse_phenomenon}). A key insight underlying these methods is that implicit-feedback datasets inherently provide a wealth of negative samples---namely, items that are skipped or not interacted with by users. This abundance of negatives facilitates multi-negative objective functions, which sharpen preference boundaries and enhance the robustness of recommendation models~\cite{DMPO,MPPO,sdpo}.

Previous research has shown that increasing negative samples in preference optimization initially benefits model performance \cite{DMPO}. However, as demonstrated in Figure~\ref{fig:dmpo_collapse}, our experimental analysis finds that beyond a critical threshold, this trend reverses: performance not only plateaus but also drops sharply, even as the training loss continues to decline.
This unexpected degradation suggests that multi-negative aggregation may introduce non-trivial negative optimization biases.
We identify this performance degradation as preference optimization collapse and attribute it, both theoretically and empirically, to gradient suppression caused by an imbalance between model‑discriminative negatives and boundary‑critical negatives (see Section~\ref{subsec:motivation}).

To mitigate this collapse, we propose \textbf{Dynamic Preference Optimization (DynamicPO)}—a plug-and-play method that maintains effective boundary learning through two adaptive mechanisms:

\begin{itemize}[label=$\bullet$, leftmargin=*]
  \item \textbf{Dynamic Boundary Negative Selection:}
  We intelligently identify the most informative boundary samples through real-time clustering, ensuring optimization focuses precisely on regions where preference discrimination remains ambiguous.
  \item \textbf{Dual-Margin Dynamic $\boldsymbol\beta$ Adjustment:}
  We assign a dynamic, sample-specific $\beta$ parameter 
  based on dual-margin discrimination difficulty, 
  adaptively concentrating gradient updates where they yield maximal boundary refinement.
\end{itemize}

Extensive experiments on three widely used recommendation datasets (i.e., Goodreads, LastFM, Steam) demonstrate the effectiveness of DynamicPO. Additionally, our plug-and-play strategy preserves training efficiency with negligible overhead, yet achieves substantial performance enhancements.
The contributions of this work can be concluded as follows:
\begin{itemize}[label=$\bullet$, leftmargin=*]
\item We experimentally identify the preference optimization collapse phenomenon and theoretically characterize it as gradient suppression caused by imbalance between model-discriminative ($\mathcal{S}$) and boundary-critical ($\mathcal{B}$) negatives
 for LLM-based recommendation systems.
\item We propose DynamicPO, a preprocessing-free and plug-and-play method that adaptively selects boundary negatives and dynamically assigns negative sample specific $\beta$ during training, thereby enabling precise preference boundary refinement and robust optimization in recommendation tasks.
\item Extensive experiments on multiple LLM backbones and datasets demonstrate that DynamicPO consistently improves recommendation performance, effectively alleviates preference optimization collapse, and generalizes across diverse multi-negative DPO objectives without incurring additional computational cost.
\end{itemize}

\section{Preliminary}
In this section, we briefly review the foundation of preference optimization and its adaptation for LLM-based recommender systems.

\subsection{Direct Preference Optimization}
Large language models (LLMs) are usually trained via supervised fine-tuning (SFT), which enhances linguistic coherence but not user alignment. Reinforcement Learning from Human Feedback (RLHF) improves alignment through reward modeling, yet training instability and cost remain high. 
DPO~\cite{DPO} provides a closed-form alternative derived from RLHF, directly aligning the model with human preference pairs without reinforcement learning:
\begin{equation}
\small
\mathcal{L}_{\text{DPO}}(\pi_{\theta}; \pi_{\text{ref}}) =
- \mathbb{E}_{(x, y_w, y_l)} \bigg[
\log \sigma \Big(
 \beta \log \frac{\pi_{\theta} (y_w \mid x)}{\pi_{\text{ref}} (y_w \mid x)} -
 \beta \log \frac{\pi_{\theta} (y_l \mid x)}{\pi_{\text{ref}} (y_l \mid x)}
\Big) \bigg].
\end{equation}
Here, $\pi_{\theta}$ denotes the optimized policy, $\pi_{\text{ref}}$ is a fixed reference model, and $\beta$ controls the regularization strength. 
DPO thus achieves stable preference alignment more efficiently than RLHF-based methods.

\subsection{Preference Optimization for LLM-based Recommenders}
Supervised fine-tuning (SFT) adapts LLMs to recommendation tasks but only maximizes the generation probability of the next item, failing to utilize negative feedback.
Direct Multi-Preference Optimization (DMPO)~\cite{DMPO} extends DPO by introducing multiple negatives to refine user preference boundaries:
\begin{equation}
\small
\mathcal{L}_{\text{DMPO}} =
-\mathbb{E}_{(x_u, y_w, y_l)} \Big[
\log \sigma
\Big(
\beta \log \frac{\pi_{\theta}(y_w|x_u)}{\pi_{\text{ref}}(y_w|x_u)}
- \frac{1}{k}\!\sum_{i=1}^{k}\!
\beta \log \frac{\pi_{\theta}(y_{l_i}|x_u)}{\pi_{\text{ref}}(y_{l_i}|x_u)}
\Big)
\Big].
\end{equation}
This formulation guides LLMs to increase the likelihood of positive items ($y_w$) while suppressing negatives $\{y_{l_i}\}_{i=1}^{k}$, enabling explicit learning of user preference boundaries and improving recommendation accuracy.

\section{Method}
\label{sec:Method}
In this section, we investigate the preference optimization collapse phenomenon in Direct Multi-Preference Optimization~\cite{DMPO} (DMPO), where model performance stagnates despite declining training loss. We attribute this collapse to gradient suppression caused by the imbalance between model-discriminative negatives and boundary-critical negatives. To resolve this, we propose DynamicPO, a dual-mechanism framework introducing: (1) adaptive boundary negative selection and (2) fine-grained $\beta$-differentiation.

\begin{figure*}[htbp]
  \centering
  \adjustbox{center}{\includegraphics[width=1.0\textwidth]{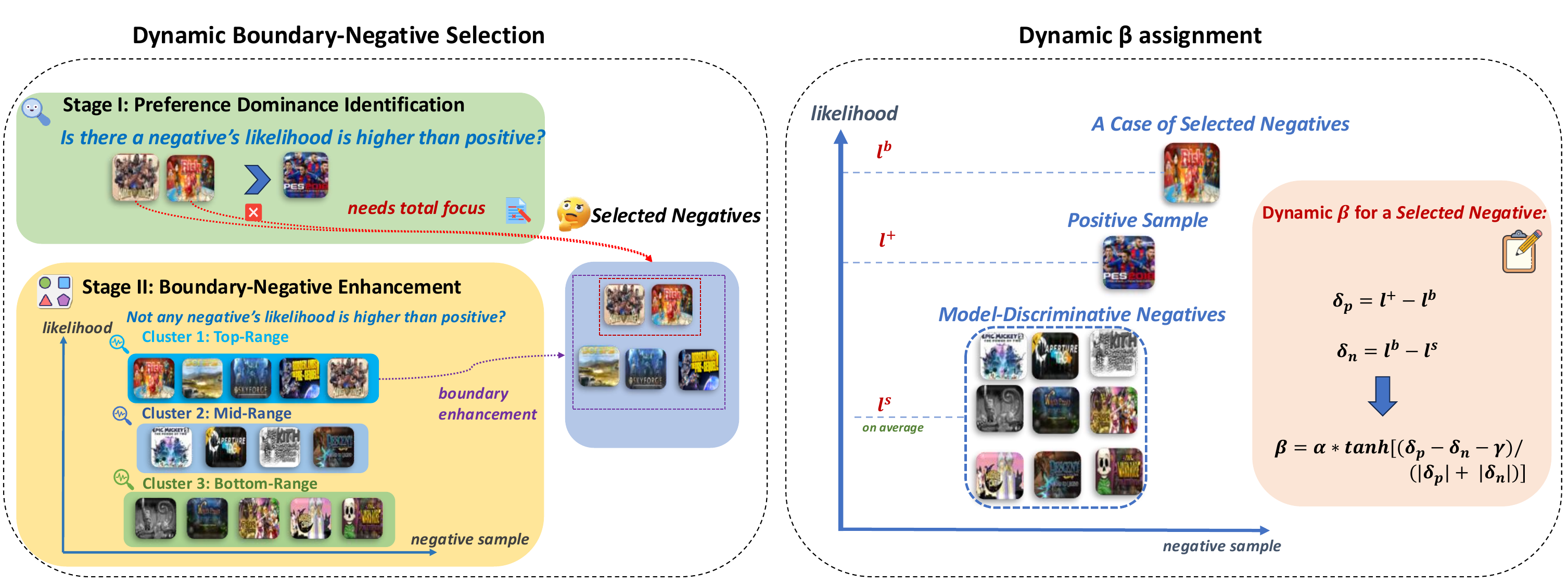}}
  \caption{Overview of DynamicPO: dynamic boundary negative selection and dynamic $\beta$-adjustment}
  \label{fig:method}
\end{figure*}

\subsection{Motivation: Gradient Suppression of Boundary-Critical Negatives in Multi-Negative DPO}
\label{subsec:motivation}
In DMPO, we observe a paradoxical phenomenon: model performance stagnates or even degrades, while training loss continues to decrease. This suggests the optimization process becomes decoupled from the true objective of refining preference boundaries.
To investigate this, we leverage the insight that the likelihood assigned to a sequence reflects the model's confidence in that response \cite{likelihood_origin}. When two responses exhibit similar likelihoods, they typically share semantic or logical coherence under the model’s internal representation\cite{likelihood,likelihood2}. 
Consequently, the likelihood gap $\Delta = \log \pi_\theta(y_w \mid x_u) - \log \pi_\theta(y_{l_i} \mid x_u)$
 serves as a quantitative measure of the model's preference resolution. By monitoring the evolution of these gaps during training, we find that negative samples can be partitioned into two distinct categories based on their proximity to the decision boundary:

\begin{itemize}[label=$\bullet$, leftmargin=*]
  \item \textbf{Model-Discriminative Negatives} ($\mathcal{S}$): negatives with large likelihood gaps ($\log \pi_\theta(y_w \mid x_u) - \log \pi_\theta(y_{l_i} \mid x_u) \gg 0$
), which the model already distinguishes well.
  \item \textbf{Boundary-Critical Negatives} ($\mathcal{B}$): negative samples on the model's decision boundary, with likelihood gaps ($\log \pi_\theta(y_w \mid x_u) - \log \pi_\theta(y_{l_i} \mid x_u) \leq 0$, 
or 
$\log \pi_\theta(y_w \mid x_u) - \log \pi_\theta(y_{l_i} \mid x_u) \approx 0$
), which indicates unresolved or ambiguous preference discrimination.
\end{itemize}

\begin{figure*}[htbp]
  \centering
  \adjustbox{center}{\includegraphics[width=1.0\textwidth]{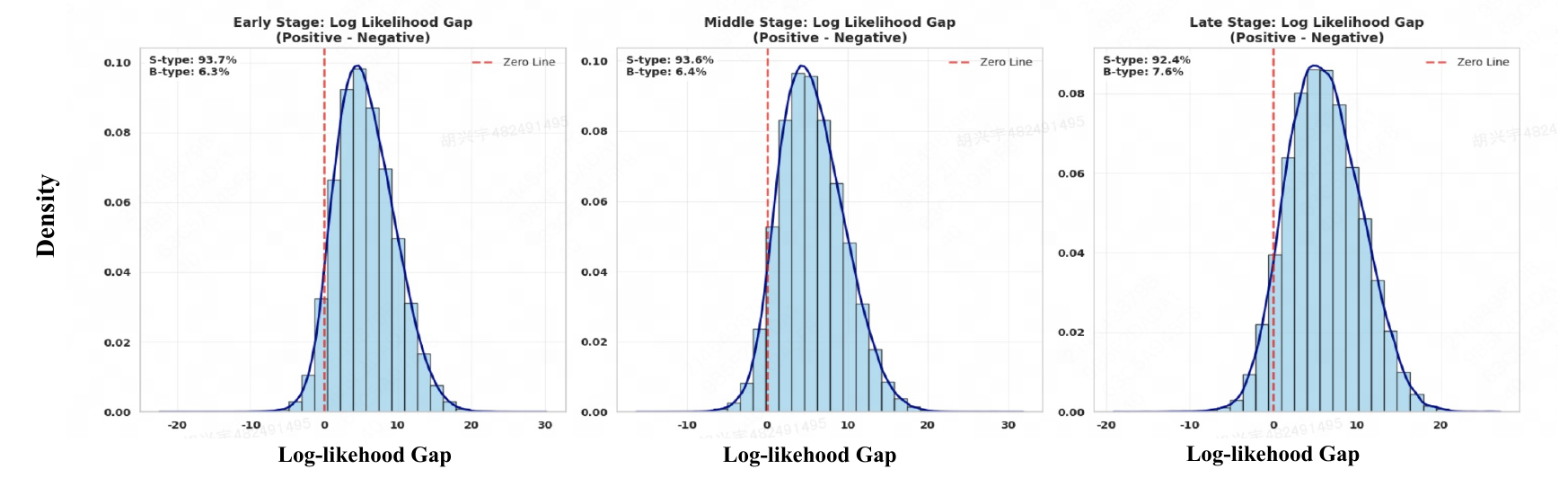}}
  \caption{Proportions of $\mathcal{S}$ and $\mathcal{B}$ through three training stages of DMPO.}
  \label{fig:S_and_B}
\end{figure*}

To empirically investigate the distribution of these two categories, we adopt zero as a coarse-grained threshold for a first-order approximation. As illustrated in Figure~\ref{fig:S_and_B}, this reveals a stark imbalance where $\mathcal{S}$ vastly outnumbers $\mathcal{B}$ throughout the training process. 
Ideally, effective optimization should resolve ambiguities, reducing the proportion of $\mathcal{B}$ over time. 
However, we observe a counter-intuitive trend: despite continuous loss reduction, the proportion of $\mathcal{B}$ actually \textbf{increases} from 6.3\% to 7.6\%. 
This signals a \textbf{boundary deterioration}: the dominance of easy negatives ($\mathcal{S}$) in the gradient creates a shortcut for loss minimization, causing the model to over-optimize trivial distinctions while neglecting—or even worsening—the resolution of critical boundary cases.

We attribute this deterioration to the \textbf{initialization bias introduced by} the SFT stage. 
Specifically, SFT equips the model with a coarse-grained discrimination capability, resulting in an abundance of already well-separated negatives ($\mathcal{S}$). 
However, standard multi-negative DPO naively aggregates these samples, inadvertently forcing the model to further amplify these trivial distinctions rather than focusing on the sparse samples near the boundary. 
Consequently, the optimization suffers from a \textbf{distributional imbalance}—where the overwhelming gradient contribution from $\mathcal{S}$ suppresses the critical updates required by the informative $\mathcal{B}$ set.
In the following, we formally analyze the mechanism by which multi-negative DPO over-optimizes model-discriminative negatives.

\subsubsection{\textbf{Mechanism of Gradient Suppression.}}

Formally, during optimization the aggregated gradient over $k$ negatives can be decomposed as:
\begin{equation}
\begin{aligned}
\frac{1}{k} \sum_{i=1}^k \nabla \log \pi_\theta(y_l^i | x_u)
&= \frac{|\mathcal{S}|}{k}
    \underbrace{\left( \frac{1}{|\mathcal{S}|} \sum_{i \in \mathcal{S}} \nabla \log \pi_\theta(y_l^i | x_u) \right)}_{\text{average gradient of $\mathcal{S}$}} \\
&\quad+ \frac{|\mathcal{B}|}{k}
    \underbrace{\left( \frac{1}{|\mathcal{B}|} \sum_{j \in \mathcal{B}} \nabla \log \pi_\theta(y_l^j | x_u) \right)}_{\text{average gradient of $\mathcal{B}$}},
\end{aligned}
\label{eq:neg_grad_decompose}
\end{equation}

Since $|\mathcal{S}|\!\gg\!|\mathcal{B}|$, the gradient is dominated by $\mathcal{S}$, forming a self-reinforcing loop where the model keeps further lowering the likelihood of already well-separated $\mathcal{S}$ negatives, even though they have been clearly distinguished from positives. 
This unnecessary amplification of easy distinctions consumes gradient capacity, leaving boundary-critical samples $\mathcal{B}$ under-optimized. 
The resulting imbalance suppresses boundary learning and drives the observed preference optimization collapse.

\subsection{Dynamic Boundary Negative Selection: Capturing Informative Samples for User Preference Modeling}
\label{subsec:method_I}

Guided by our theoretical insights, we prioritize \textbf{Boundary-Critical Negatives} ($\mathcal{B}$) as the primary source of meaningful optimization signals.
Standard DMPO can fall into a deceptive convergence, whereby the loss decreases even as preference resolution deteriorates. This occurs because the model focuses on over-optimizing trivial gaps. We mitigate this through two core principles:
\begin{itemize}[label=$\bullet$, leftmargin=*]
    \item \textbf{Principle 1:}  Focus on boundary negatives resembling positives, as they best capture unresolved preferences and refine the decision boundary.
    \item \textbf{Principle 2:} Select boundary negatives adaptively according to sample informativeness, avoiding fixed thresholds.

\end{itemize}

These principles are operationalized via a \textbf{fully online and preprocessing-free} selection strategy consisting of two hierarchical stages:

\textbf{Stage 1: Preference Dominance Identification.} We first identify critical preference violations where the model's current policy contradicts the intended ordering. Let $L_p$ and $L(n)$ denote the log-likelihoods of the positive and negative samples $n \in \mathcal{N}$, respectively. We define the violation set as:
\begin{equation}
\mathcal{N}_{vio} = \{ n \in \mathcal{N} \mid L(n) > L_p \}
\end{equation}
If $\mathcal{N}_{vio} \neq \emptyset$, these samples directly expose boundary deficiencies and are prioritized for optimization by setting the selected boundary negative set $\mathcal{B} = \mathcal{N}_{vio}$.

\textbf{Stage 2: Boundary Negative Enhancement via Likelihood Clustering.} If no such violation exists ($\mathcal{N}_{vio} = \emptyset$), we focus on negatives near the decision boundary to resolve preference ambiguities. We adaptively partition the negative samples into three clusters $\{\mathcal{C}_{top}, \mathcal{C}_{mid}, \mathcal{C}_{bot}\}$ using $k$-means clustering ($k=3$) based on their log-likelihoods $\{L(n)\}_{n \in \mathcal{N}}$. The selection is then defined as $\mathcal{B} = \mathcal{C}_{top}$, where $\mathcal{C}_{top}$ denotes the cluster with the highest centroid. This adaptive grouping allows the model to flexibly capture informative boundary negatives across varying likelihood distributions without relying on rigid static thresholds.

By adaptively balancing major violation correction with boundary refinement, this mechanism effectively mitigates signal dilution and enhances the precision of preference modeling.

\subsection{Dynamic \texorpdfstring{$\beta$}{beta}-Adjustment: Fine-Grained Boundary Regularization}
In DPO-based frameworks, the hyperparameter $\beta$ modulates the Kullback-Leibler (KL) regularization between the learned policy $\pi_\theta$ and the reference policy $\pi_{\mathrm{ref}}$ \cite{DPO}. A smaller $\beta$ facilitates aggressive adaptation to preference data (plasticity), whereas a larger $\beta$ enforces conservative alignment to preserve the reference model's prior knowledge (stability).
While $\beta$-DPO~\cite{betadpo} introduces a dynamic $\beta$ for natural language generation, we observe that its direct application to our \textit{Dynamic Boundary Negative Selection} unexpectedly yields a 0.32\% performance drop. This suggests that existing adjustment strategies—primarily designed for open-ended generation—fail to account for the \textit{multi-negative} and \textit{implicit-feedback} nature of recommendation tasks.

To this end, we propose a fine-grained dynamic $\beta$-adjustment strategy to adaptively amplify optimization signals from each selected boundary negative and prevent boundary dilution through two principles:
\begin{itemize}[label=$\bullet$, leftmargin=*]
    \item \textbf{Principle 1:} Assign $\beta$ dynamically to each boundary negative based on its own informativeness.
    \item \textbf{Principle 2:} Modulate $\beta$ using both positive–negative ambiguity and distance from easy negatives to emphasize harder samples.
\end{itemize}

To operationalize these principles, we introduce a dual-margin mechanism for each selected boundary negative sample:

\begin{itemize}[label=$\bullet$, leftmargin=*]
    \item \textbf{Positive-to-Boundary Margin} ($\delta_p = L^+ - L^b$): measures \textit{boundary ambiguity} using the likelihoods of the positive sample ($L^+$) and the boundary negative ($L^b$).
    \item \textbf{Boundary-to-Easy Margin} ($\delta_n = L^b - L^e$): quantifies \textit{informativeness contrast} relative to the average log-likelihood of normal negatives ($L^e$).
\end{itemize}

We then compute the dynamic $\beta$ for each sample as follows:
\begin{equation}
\beta =\beta_0\cdot(1+ \alpha \cdot \tanh\left( \frac{\delta_p - \delta_n - \gamma}{|\delta_p| + |\delta_n|} \right)),
\end{equation}

In this formula, $\beta_0$ is the base regularization weight and $\alpha$ scales the adjustment intensity. Crucially, $\gamma$ denotes the \textit{intrinsic preference margin}, representing the natural likelihood superiority that a positive sample is expected to maintain over a negative one. The normalized denominator $(|\delta_p| + |\delta_n|)$ ensures the adjustment is sensitive to the \textit{relative} likelihood distribution. Notably, the $\tanh(\cdot)$ function constrains $\beta$ within a stable range, yielding updates that are both expressive for challenging cases and robust against outliers. 

\textbf{The intuition behind this design is to adaptively balance plasticity and stability based on sample-specific difficulty.} Specifically, for a challenging and informative negative that exhibits high ambiguity (small $\delta_p$) and significant contrast from discriminative negatives (large $\delta_n$), the numerator $(\delta_p - \delta_n - \gamma)$ becomes negative. This leads to a decreased $\beta$, thereby reducing KL-regularization and allowing the model more \textit{plasticity} to aggressively refine the decision boundary. Conversely, for trivial samples that are easily discriminable, $\beta$ increases to enforce \textit{stability} and prevent the model from over-adapting to non-critical signals.
By precisely customizing $\beta$, DynamicPO provides fine-grained gradients that focus the model's capacity on refining the most critical preference regions.

\section{Experiments}

In this section, we conduct extensive experiments to evaluate the effectiveness and robustness of \textbf{DynamicPO}. We first outline the experimental setup, encompassing diverse LLM backbones, benchmark datasets, and representative baselines. Subsequently, we present a comparative analysis of experimental results, focusing on DynamicPO's ability to mitigate preference optimization collapse and its generalization across various multi-negative objectives. Finally, we provide in-depth analyses of reward boundary evolution and computational efficiency to further validate the superiority of our approach.

\subsection{Experimental Settings}
\subsubsection{Base Model}
Our approach is evaluated using three distinct base models: Llama2-7b-hf for the main experiments, supplemented by Llama3-8B-Instruct and Qwen2.5-7B-Instruct for exploratory studies. The diversity of these base models provides a robust foundation for our experimental analysis.

\subsubsection{Datasets}
We evaluate our approach on three widely used recommendation datasets from diverse domains: LastFM~\cite{lastfm} (music recommendation), Goodreads\footnote{\url{https://www.goodreads.com}} (book ratings and reviews), and Steam\footnote{\url{https://store.steampowered.com/}} (video game ratings and reviews). We follow prior work~\cite{liao2024llara} for preprocessing and chronological 8:1:1 splitting. Detailed statistics for each dataset are summarized in Table~\ref{table_data_characteristics}.

\subsubsection{Evaluation}
Following prior work~\cite{liao2024llara,sdpo}, we adopt HitRatio@1 to evaluate recommendation accuracy via re-ranking: for each sequence, models identify the correct item from a candidate set containing 20 negative samples plus the ground-truth item. Valid Ratio measures LLM-specific behaviors (e.g., instruction-following), and consistently approaches 1.0 (>0.95) across all experiments—indicating near-perfect instruction compliance. Consequently, we omit Valid Ratio in exploration studies as only HitRatio@1 reflects recommendation capability.

\begin{table}[ht]
    \centering
    \caption{Statistics of Datasets.}
    \begin{tabular}{l@{\hspace{20pt}}c@{\hspace{20pt}}c@{\hspace{20pt}}c}
        \hline
        Dataset & LastFM & Steam & Goodreads \\
        \hline
        \# Sequence & 1,220 & 11,938 & 6,031 \\
        \# Item & 4,606 & 3,581 & 4,500 \\
        \# Interaction & 73,510 & 274,726 & 220,100 \\
        \hline
    \end{tabular}
    \label{table_data_characteristics}
\end{table}

\subsubsection{Baselines}
We compare our approach with several representative methods, covering both traditional and LLM-based paradigms. Traditional baselines include GRU4Rec~\cite{GRU4rec} (GRU for sequential modeling), Caser~\cite{Caser} (convolutional filters for sequence patterns), and SASRec~\cite{SASRec} (self-attention for long-term dependencies). LLM-based models include MoRec~\cite{MoRec} (BERT for text encoding and SASRec for sequences), TALLRec~\cite{bao2023tallrec} (instruction-tuned LLMs for recommendation), LLaRA~\cite{liao2024llara} (leveraging both language and traditional sequential encodings), and DMPO~\cite{DMPO} (multi-negative preference optimization for recommendation).

\subsubsection{Implementation Details}
We implement our experiments on 4 NVIDIA A100 GPUs.
For LLM-based recommenders, supervised fine-tuning is performed for up to 5 epochs. Preference optimization-based approaches are further refined through a preference alignment phase, spanning 3 additional epochs. Following the setup in previous work~\cite{sdpo}, the hyperparameter $\beta$ is set to 1, and 15 negative samples are incorporated during preference optimization. For DynamicPO’s dynamic-$\beta$ adjustment, we set $\alpha=0.5$ and $\gamma=6$. Refer to our code repository for full implementation details.

\subsection{Experiment Results}

\begin{table}[t]
\caption{The performance comparison on three real-world datasets.}
\centering
\small
\begin{adjustbox}{max width=\textwidth}
\begin{tabular}{llccccccc}
\toprule
 &  & \multicolumn{2}{c}{LastFM} & \multicolumn{2}{c}{Goodreads} & \multicolumn{2}{c}{Steam} \\
\cmidrule(lr){3-4}\cmidrule(lr){5-6}\cmidrule(lr){7-8}
\textbf{Category} & \textbf{Method} & HitRatio@1 & ValidRatio & HitRatio@1 & ValidRatio & HitRatio@1 & ValidRatio\\
\midrule
\multirow{3}{*}{\textbf{Traditional}} &
GRU4Rec\cite{GRU4rec} & 0.2616 & 1.0000 & 0.3867 & 1.0000 & 0.4168 & 1.0000\\
& Caser\cite{Caser} & 0.2233 & 1.0000 & 0.4174 & 1.0000 & 0.4368 & 1.0000\\
& SASRec\cite{SASRec} & 0.2233 & 1.0000 & 0.3581 & 1.0000 & 0.4010 & 1.0000\\
\midrule
\multirow{3}{*}{\textbf{LLM-based}} &
MoRec\cite{MoRec} & 0.1652 & 1.0000 & 0.2877 & 1.0000 & 0.3911 & 1.0000\\
& TALLRec\cite{bao2023tallrec} & 0.4180 & 0.9836 & 0.4983 & 0.9573 & 0.4637 & 0.9840\\
& LLaRA\cite{liao2024llara} & 0.4508 & 0.9918 & 0.5292 & 0.9950 & 0.4949 & 0.9975\\
\midrule
\multirow{2}{*}{\textbf{PO-based}}
& DMPO\cite{DMPO}         & 0.5848 & 0.9924 & 0.5349 & 0.9717 & 0.6383 & 0.9704 \\
& \cellcolor{gray!8}\textbf{DynamicPO} 
  & \cellcolor{gray!8}\textbf{0.6661} 
  & \cellcolor{gray!8}0.9980 
  & \cellcolor{gray!8}\textbf{0.6728} 
  & \cellcolor{gray!8}0.9900 
  & \cellcolor{gray!8}\textbf{0.6990} 
  & \cellcolor{gray!8}0.9789 \\
\bottomrule
\end{tabular}
\end{adjustbox}
\end{table}

\begin{figure}[htbp]
  \centering
  \subfloat[HitRatio@1 for DMPO and DynamicPO with varying negative sample scales\label{fig:dmpo_dynamic_neg}]{
    \includegraphics[width=0.48\textwidth]{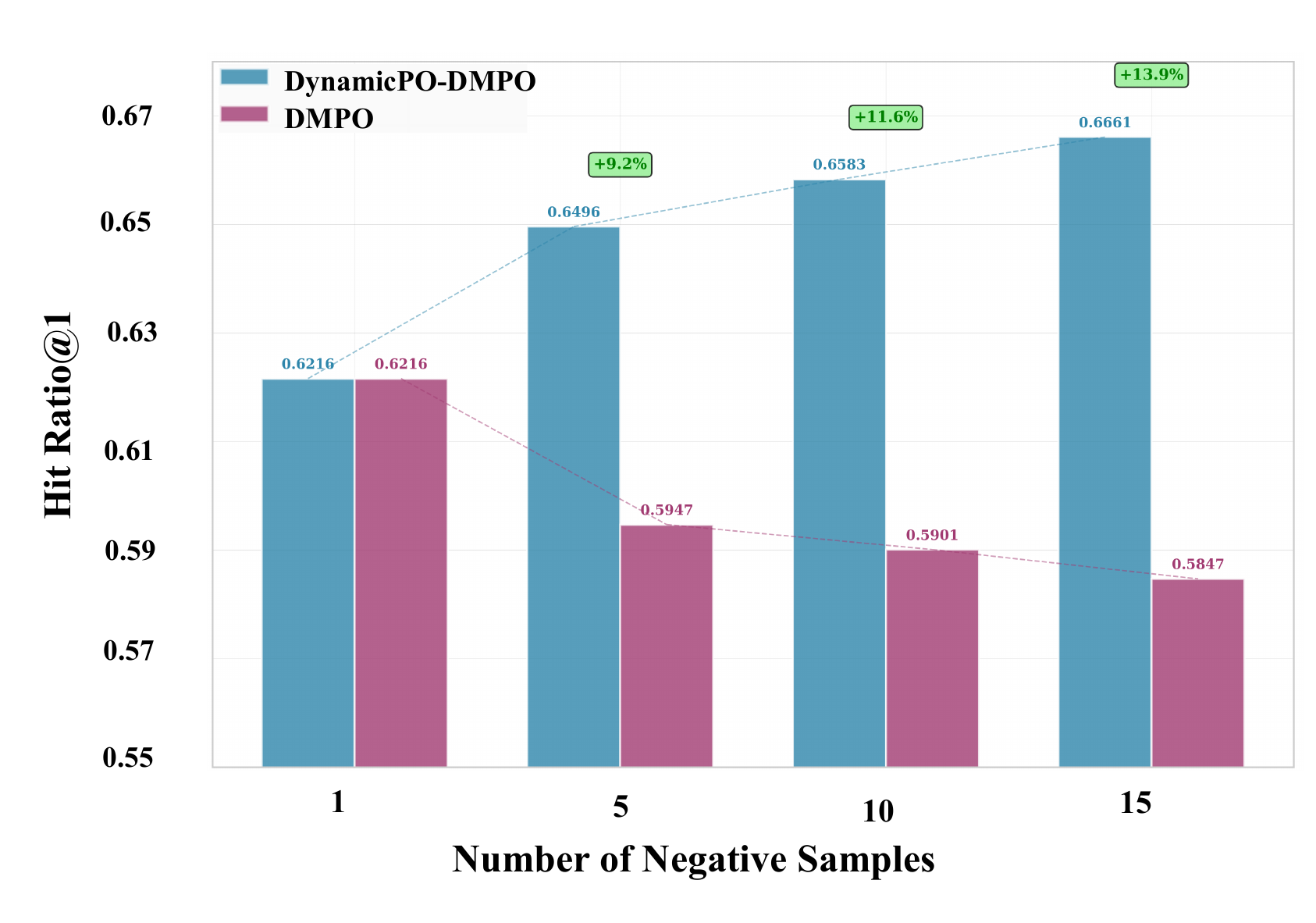}
  }\hfill
  \subfloat[Reward accuracy evolution (20\% intervals)\label{fig:reward_accuracy}]{
    \includegraphics[width=0.48\textwidth]{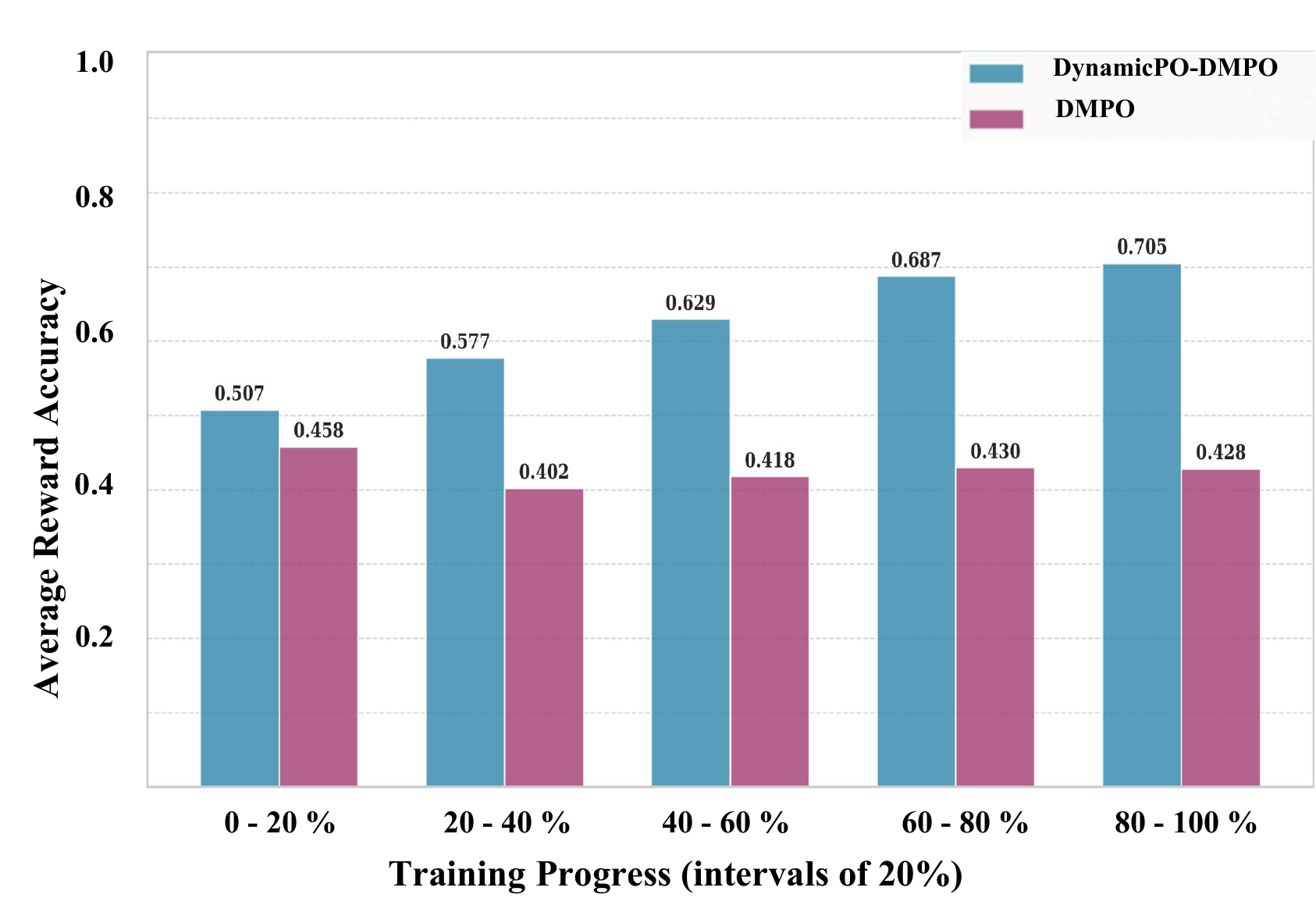}
  }
  \caption{Effect of negative sample scaling and reward evolution on model performance in DMPO and DynamicPO}
\end{figure}

\textbf{DynamicPO averts the preference optimization collapse.}
We integrate our proposed mechanisms into DMPO to form DynamicPO-DMPO (simply DynamicPO hereafter). As shown in Figure~\ref{fig:dmpo_dynamic_neg}, increasing the number of negative samples leads to a consistent decline in DMPO's performance. In contrast, DynamicPO maintains a consistently increasing performance as the number of negatives increases. 
Notably, with 15 negatives, DynamicPO boosts HitRatio@1 from 58.47\% to 66.61\% on \textbf{Llama2-7b-hf}.
To investigate cross-model robustness, we evaluate two additional backbones (Table~\ref{tab:DMPO_DynamicPO_Llama3_Qwen2.5}). 
\textbf{Llama3-8B-Instruct} achieves +17.6\% on LastFM and +15.0\% on Goodreads, while \textbf{Qwen2.5-7B-Instruct} yields +9.2\% and +11.2\%, respectively.
These consistent gains across diverse LLM backbones validate DynamicPO’s effectiveness in refining preference boundaries and enhancing recommendation performance.

\begin{table}[t]
\centering
\caption{Performance comparison of DMPO and DynamicPO on Llama3-8B-Instruct and Qwen2.5-7B-Instruct across LastFM and Goodreads datasets.}
\label{tab:DMPO_DynamicPO_Llama3_Qwen2.5}
\small
\begin{adjustbox}{max width=\textwidth}
\begin{tabular}{llcc}
\toprule
\multirow{2}{*}{\textbf{Model}} & \multirow{2}{*}{\textbf{Method}} & \multicolumn{2}{c}{\textbf{Dataset}} \\
\cmidrule(lr){3-4}
 &  & \textbf{LastFM} & \textbf{Goodreads} \\
\midrule
\multirow{2}{*}{\textbf{Llama3-8B-Instruct}} 
& DMPO & 0.6232 & 0.6645 \\
& \cellcolor{gray!8}\textbf{DynamicPO} 
  & \cellcolor{gray!8}\textbf{0.7331} 
  & \cellcolor{gray!8}\textbf{0.7641} \\
\midrule
\multirow{2}{*}{\textbf{Qwen2.5-7B-Instruct}} 
& DMPO & 0.5892 & 0.6617 \\
& \cellcolor{gray!8}\textbf{DynamicPO} 
  & \cellcolor{gray!8}\textbf{0.6433} 
  & \cellcolor{gray!8}\textbf{0.7359} \\
\bottomrule
\end{tabular}
\end{adjustbox}
\end{table}

\textbf{DynamicPO demonstrates strong generalization across diverse multi-negative objectives.}
Beyond addressing the preference optimization collapse specific to DMPO, we seek to validate the universality of DynamicPO.  We therefore extend our proposed mechanisms to other representative multi-negative DPO methods, such as MPPO~\cite{MPPO} and S-DPO~\cite{sdpo}. 
As illustrated in Table~\ref{tab:MPPO_SDPO_performance},  DynamicPO consistently outperforms the original multi-negative DPO methods on all datasets. For instance, on the LastFM dataset, it enhances the HitRatio@1 of MPPO from 0.6597 to 0.6906 and S-DPO from 0.6617 to 0.6666. Similar performance gains are observed on Goodreads and Steam, confirming that DynamicPO serves as a robust plug-and-play solution for refining preference boundaries across various multi-negative optimization objectives.

\begin{table}[t]
\caption{HitRatio@1 of DynamicPO across other representative multi-preference objectives.}
\label{tab:MPPO_SDPO_performance}
\centering
\small
\begin{adjustbox}{max width=\textwidth}
\begin{tabular}{llccc}
\toprule
&  & \textbf{LastFM} & \textbf{Goodreads} & \textbf{Steam} \\
\cmidrule(lr){3-3}\cmidrule(lr){4-4}\cmidrule(lr){5-5}
\textbf{Category} & \textbf{Method} & HitRatio@1 & HitRatio@1 & HitRatio@1 \\
\midrule
\multirow{2}{*}{\textbf{MPPO\cite{MPPO}}} 
& naive & 0.6597 & 0.6993 & 0.7614 \\
& \cellcolor{gray!8}\textbf{DynamicPO} 
  & \cellcolor{gray!8}\textbf{0.6906} 
  & \cellcolor{gray!8}\textbf{0.7226} 
  & \cellcolor{gray!8}\textbf{0.8069} \\
\midrule
  \multirow{2}{*}{\textbf{S-DPO\cite{sdpo}}} 
& naive & 0.6617 & 0.6778 & 0.6948 \\
& \cellcolor{gray!8}\textbf{DynamicPO} 
  & \cellcolor{gray!8}\textbf{0.6666} 
  & \cellcolor{gray!8}\textbf{0.6843} 
  & \cellcolor{gray!8}\textbf{0.6998} \\
\bottomrule
\end{tabular}
\end{adjustbox}
\end{table}

 \textbf{DynamicPO forges a refined and robust preference boundary.} To evaluate the quality of the learned boundaries, we track the ``reward win rate,'' defined as the frequency with which a positive sample's reward surpasses those of all negative samples.
Figure~\ref{fig:reward_accuracy} shows that DynamicPO significantly outperforms DMPO; in the final training stages, the win rate climbs from 42.8\% to 70.5\% (+27.7\%). 
This substantial gain underscores DynamicPO's enhanced discriminative power and its capacity to forge sharper boundaries compared to standard preference optimization techniques.

 \textbf{DynamicPO incurs negligible computational overhead.}
To assess the computational efficiency of DynamicPO, we measure its training duration when integrated with DMPO across three base models. 
As shown in Table~\ref{tab:A100_GPU_time}, DynamicPO incurs merely 0.85\% additional training time versus standard DMPO. 
These results demonstrate that our method achieves superior performance without substantially increasing GPU resource consumption.

\begin{table}[t]
\caption{A100 GPU time consumption for training DMPO and DynamicPO across all base models (15 negative samples).}
\label{tab:A100_GPU_time}
\centering
\small
\begin{adjustbox}{max width=\textwidth}
\begin{tabular}{lcc}
\toprule
\textbf{Base Model} & \textbf{DMPO} & \textbf{DynamicPO} \\
\midrule
Llama2-7b-hf        & 4$\cdot$A100$\times$16h38min  & 4$\cdot$A100$\times$16h41min (+3min) \\
Llama3-8B-Instruct  & 4$\cdot$A100$\times$15h29min  & 4$\cdot$A100$\times$15h42min (+13min) \\
Qwen2.5-7B-Instruct & 4$\cdot$A100$\times$14h49min  & 4$\cdot$A100$\times$14h57min (+8min) \\
\midrule
Avg.~GPU time & 62.58h$\cdot$A100 & 63.11h$\cdot$A100 (+0.85\%) \\
\bottomrule
\end{tabular}
\end{adjustbox}
\end{table}

\subsection{Study of DynamicPO}
\subsubsection{Ablation Study}
\label{subsubsec:ablation_study}
To evaluate the contribution of each component in DynamicPO, we conduct ablation studies by systematically removing Stage~I (preference dominance identification), Stage~II (boundary signal enhancement), both stages, and the dynamic $\beta$ adjustment. Experiments on LastFM, Goodreads, and Steam are summarized in Table~\ref{tab:ablation}. 
Results show that eliminating Stage~II yields the largest performance drop (e.g., HitRatio@1 on LastFM declines from 0.6661 to 0.6549), confirming its key role in refining decision boundaries. Removing Stage~I or dynamic $\beta$ adjustment causes moderate declines (to 0.6621 and 0.6593, respectively), while omitting both stages severely degrades performance (0.5884 on LastFM). 
Overall, these results demonstrate that the staged negative selection and adaptive $\beta$ mechanisms in DynamicPO are essential for robust and generalizable preference optimization.

\begin{table}[t]
\caption{Ablation studies of DynamicPO on three datasets.}
\label{tab:ablation}
\centering
\small
\begin{adjustbox}{max width=\textwidth}
\begin{tabular}{lccc}
\toprule
\textbf{Method} & \textbf{LastFM} & \textbf{Goodreads} & \textbf{Steam} \\
\midrule
DMPO & 0.5848 & 0.5349 & 0.6383 \\
\cellcolor{gray!8}\textbf{DynamicPO} 
  & \cellcolor{gray!8}\textbf{0.6661} 
  & \cellcolor{gray!8}\textbf{0.6728} 
  & \cellcolor{gray!8}\textbf{0.6990} \\
\hspace{2em}w/o Stage (I)       & 0.6621 & 0.6644 & 0.6914 \\
\hspace{2em}w/o Stage (II)      & 0.6549 & 0.6594 & 0.6830 \\
\hspace{2em}w/o Stage (I \& II) & 0.5884 & 0.5365 & 0.6383 \\
\hspace{2em}w/o dynamic $\beta$ & 0.6593 & 0.6678 & 0.6913 \\
\bottomrule
\end{tabular}
\end{adjustbox}
\end{table}

\subsubsection{Exploration of Selection Strategies: Adaptive versus Rigid Selection}
Following our theoretical analysis of preference optimization collapse, we initially considered a Top-$K$ strategy that selects the $K$ negatives with the highest likelihood as a potential remedy. To conduct a fair comparison between this candidate and our proposed adaptive selection, we tracked the training process of DynamicPO and observed that it selects an average of 3.4 negatives per instance. Accordingly, we set $K \in \{2, 3, 4\}$ for the Top-$K$ baseline to bracket this average, ensuring equivalent signal density across methods.
As shown in Table~\ref{tab:topk_dynamicpo}, DynamicPO consistently outperforms all Top-$K$ variants, exhibiting a maximum margin of 3.2\% on LastFM. This performance gap reveals the limitations of rigid truncation: fixed thresholds lack distributional awareness and often exclude ambiguous yet informative samples residing near the decision boundary. In contrast, DynamicPO adaptively identifies these boundary zones, capturing hard negatives that precisely match the model's current discriminative capacity, thereby enabling more robust preference learning.

\begin{table}[ht]
    \centering
    \caption{Performance improvements of DynamicPO over traditional Top-K baselines on LastFM and Goodreads}
    \begin{tabular}{llcc}
        \hline
         Method &Strategy & LastFM & Goodreads \\
        \hline
        \multirow{3}{*}{Top-K}
            & k=2 & 0.6452 & 0.6561 \\
            & k=3 & 0.6492 & 0.6645 \\
            & k=4 & 0.6476 & 0.6594 \\
        \hline
        \multirow{1}{*}{\cellcolor{gray!8}\textbf{DynamicPO}}
            & \cellcolor{gray!8}\textbf{DynamicPO} & \cellcolor{gray!8}\textbf{0.6661} & \cellcolor{gray!8}\textbf{0.6728} \\
        \hline
    \end{tabular}
    \label{tab:topk_dynamicpo}
\end{table}

\subsubsection{Hyperparameter Analysis}
\begin{figure}[htbp]
  \centering
  \subfloat[Study of $\gamma$ on HitRatio@1\label{fig:gamma_beta}]{
    \includegraphics[width=0.48\textwidth]{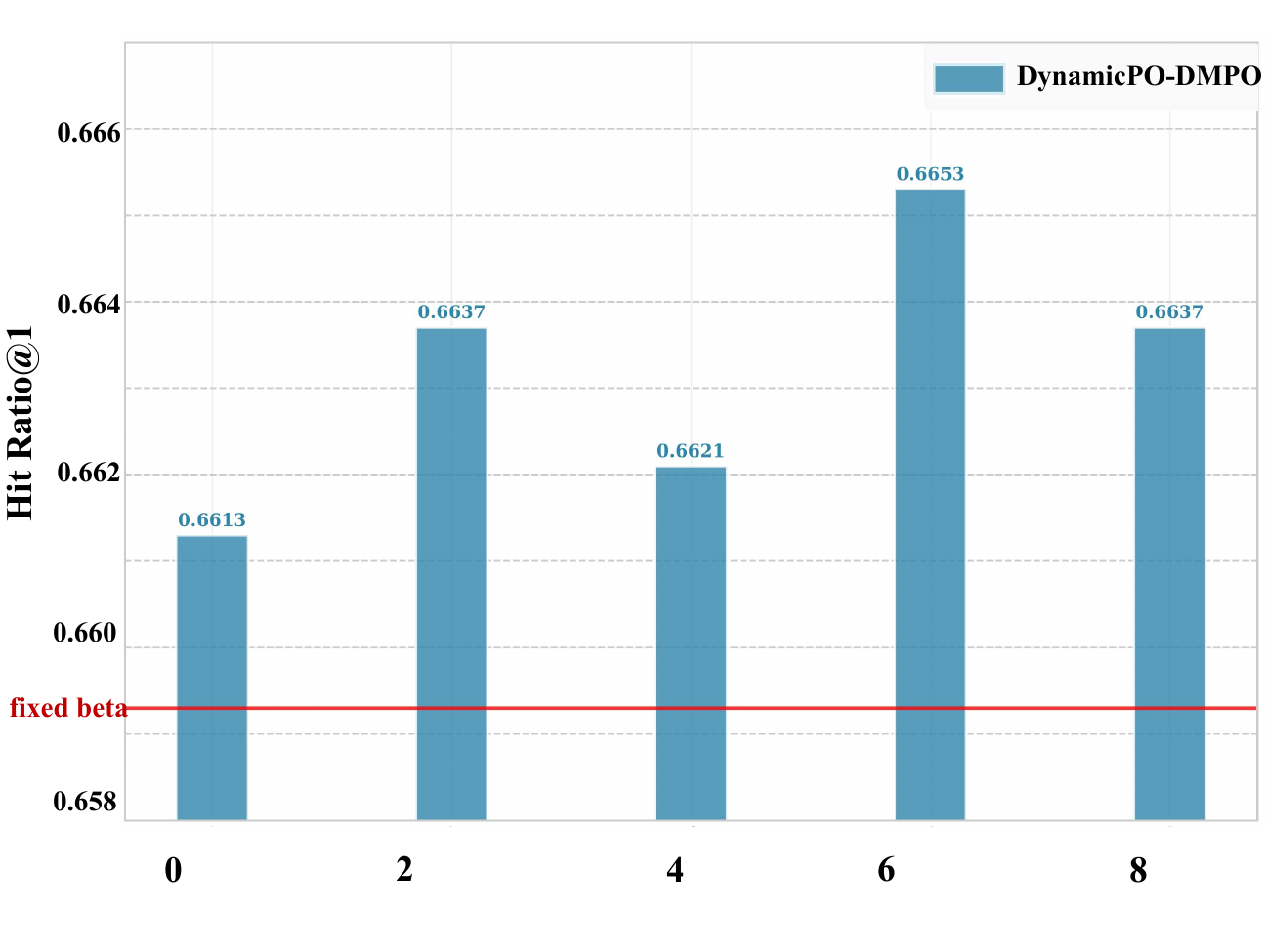}
  }\hfill
  \subfloat[Study of $\alpha$ on HitRatio@1\label{fig:alpha_beta}]{
    \includegraphics[width=0.48\textwidth]{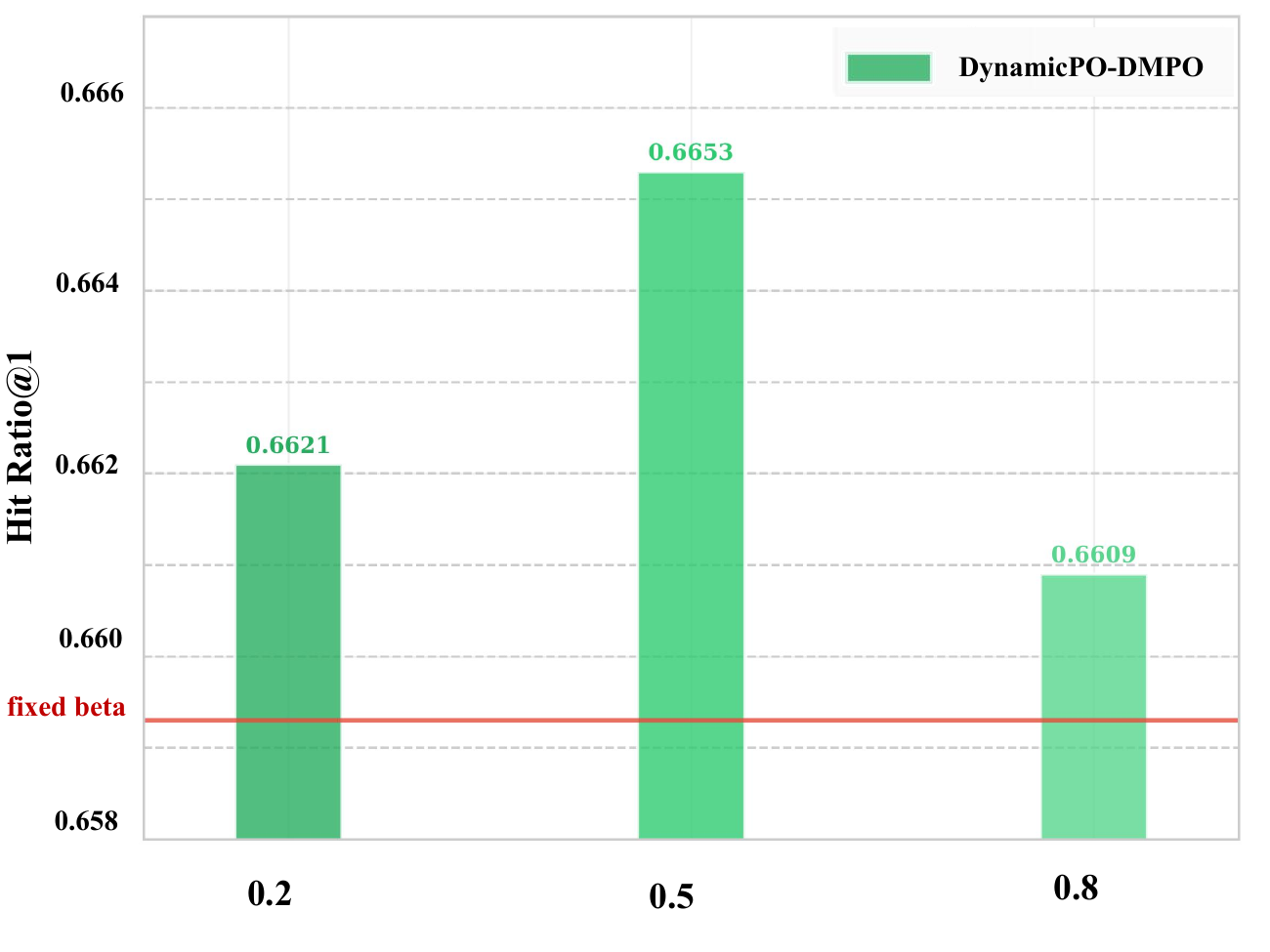}
  }
  \caption{Study of $\gamma$ and $\alpha$ of DynamicPO on LastFM.}
  \label{fig:hyperparameter}
\end{figure}

We investigate the sensitivity of DynamicPO to two key hyperparameters in the dynamic $\beta$ formula: $\gamma$ (the \textit{intrinsic preference margin}) and $\alpha$ (the factor scaling the \textit{adjustment intensity}). As illustrated in Figure~\ref{fig:hyperparameter}, DynamicPO maintains high and stable performance across a broad range of values for both parameters, consistently exceeding the fixed $\beta$ baseline. Specifically, the HitRatio@1 remains relatively constant despite variations in $\gamma$ and $\alpha$, suggesting that the model is not overly sensitive to hyperparameter tuning. This stability underscores the robustness and practical reliability of DynamicPO in diverse recommendation scenarios.

\section{Related Work}
\textbf{LLM for Recommendation.} Sequential recommendation helps users manage information overload by mining interests from their past behaviors. The emergence of LLMs, with strong generative and reasoning abilities, has driven new advances in recommendation systems (LLM4Rec~\cite{charrec,liao2024llara,bao2023tallrec,wang2024rethinking}), where recommendation is reformulated as a language modeling task for multitask learning and zero-shot generalization.
Recent work applies LLMs in two main ways: (1) \textbf{LLMs as Recommenders}, directly generating items from user histories~\cite{yang2023large,geng2022recommendation,DPO,sdpo}; and (2) \textbf{LLMs as Enhancers}, utilizing textual features to enhance traditional pipelines~\cite{houyupeng,MoRec}. Recently, exploring item representation during fine-tuning (e.g., integrating collaborative signals~\cite{zhang2025collm,kong2024customizing}, adjusting numeric representations) has further boosted performance. 

\noindent\textbf{Preference Optimization in LLMs.}
Preference optimization methods for recommendation, such as DMPO~\cite{DMPO}, S-DPO~\cite{sdpo}, and MPPO~\cite{MPPO}, introduce multi-negative DPO methods to better capture user interests. Although increasing negative samples can boost performance, our experiments and theoretical analysis reveal an unexpected issue—\emph{preference optimization collapse}.
To address this, we propose DynamicPO, a plug-and-play approach that mitigates collapse and improves recommendation performance.

\section{Conclusion}

In this work, we identify preference optimization collapse in LLM-based recommender systems, theoretically analyze its underlying causes, and based on this analysis, we propose Dynamic Preference Optimization (DynamicPO)—a plug-and-play method to address this challenge.
DynamicPO introduces two adaptive mechanisms: dynamic boundary negative selection via real-time clustering and dual-margin dynamic $\beta$ adjustment.
By prioritizing boundary-critical negatives through real-time clustering and customizing optimization strength for each negative sample, DynamicPO ensures effective refinement of preference boundaries and robust user interest modeling. Our approach is efficient, introducing negligible computational overhead, and can be seamlessly integrated into existing multi-negative preference optimization objectives. Extensive experiments across diverse datasets and backbone models demonstrate that DynamicPO prevents optimization collapse while consistently improving performance on LLM-based sequential recommendation. These results highlight the importance of boundary-aware dynamic optimization for robust and efficient preference alignment in LLM-based recommendation. Future work will investigate integrating structural information, such as user relationship networks, into dynamic preference optimization to further enhance explainability and adaptability.

\subsubsection*{Acknowledgements.}
This work was supported by the computing resources provided by Meituan.

\appendix
\section{Additional Experimental Results} \label{app:additional_experiments}

\subsection{Negative-Set Scaling across Multi-Negative Objectives}
\label{app:negative_scaling}

The main experiments reveal preference optimization collapse in DMPO, where recommendation performance deteriorates once the negative-set size exceeds an effective range despite the continued reduction in training loss. To examine whether this phenomenon generalizes beyond DMPO, we conduct supplementary scaling experiments on DMPO, MPPO, and S-DPO using LastFM. Compared with the main experiments, we further extend the maximum number of negatives from 15 to 19 and report the complete vanilla scaling curves of all three multi-negative DPO objectives. We then evaluate DynamicPO under the largest negative-set size of 19 to examine whether it can effectively exploit enlarged negative sets across different objective formulations. All supplementary scaling experiments are conducted on NVIDIA H200 GPUs. We also reproduce the LastFM experiments from the main paper under the same H200 setting. The checkpoints are publicly available on \href{https://huggingface.co/xingyuHuxingyu/DynamicPO}{Hugging Face}.

\begin{figure}
\centering
\includegraphics[width=0.68\linewidth]{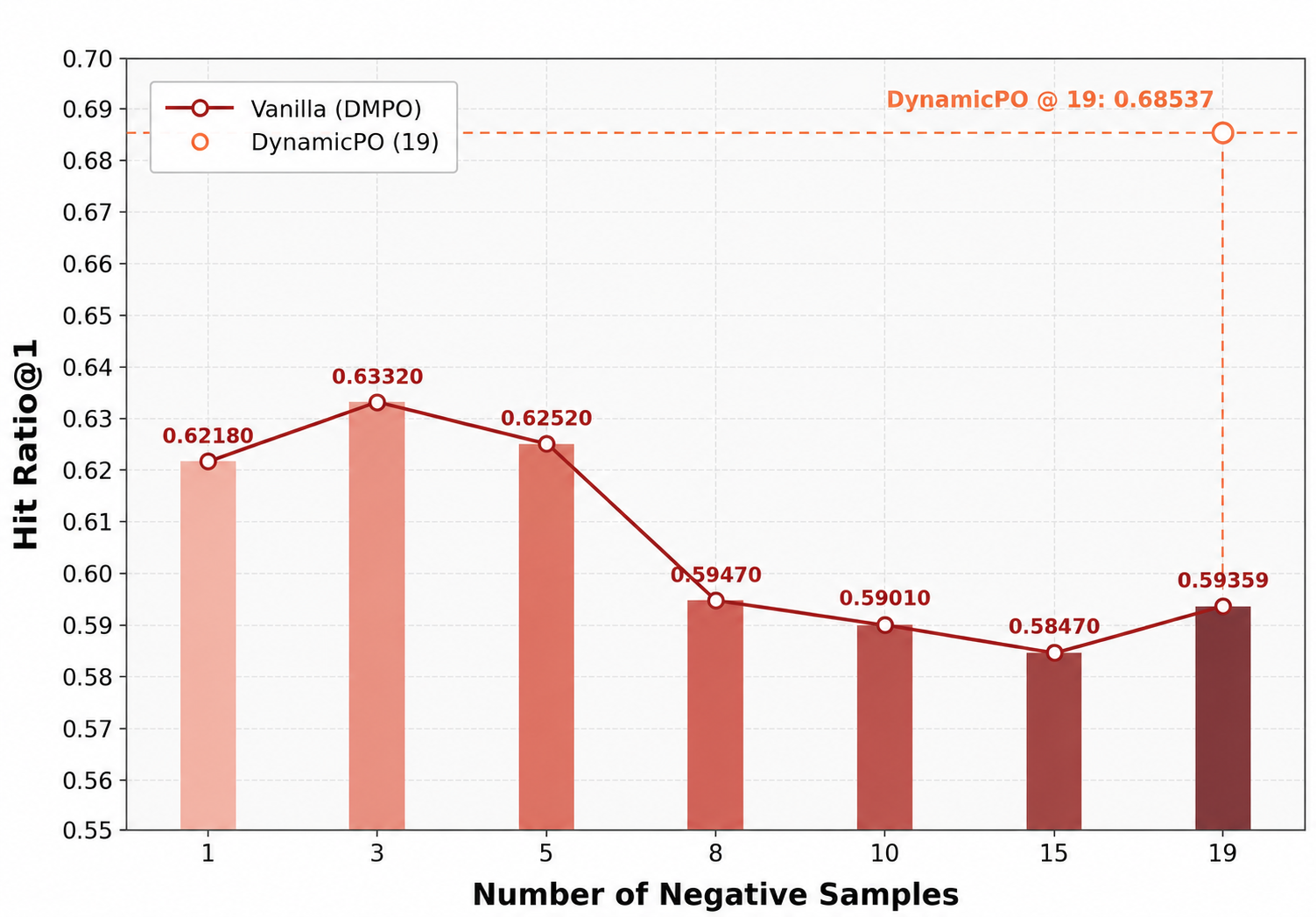}
\caption{Scaling behavior of DMPO as the number of negative samples increases.}
\label{fig:dmpo_negative_scaling}
\end{figure}

\begin{figure}
\centering
\includegraphics[width=0.68\linewidth]{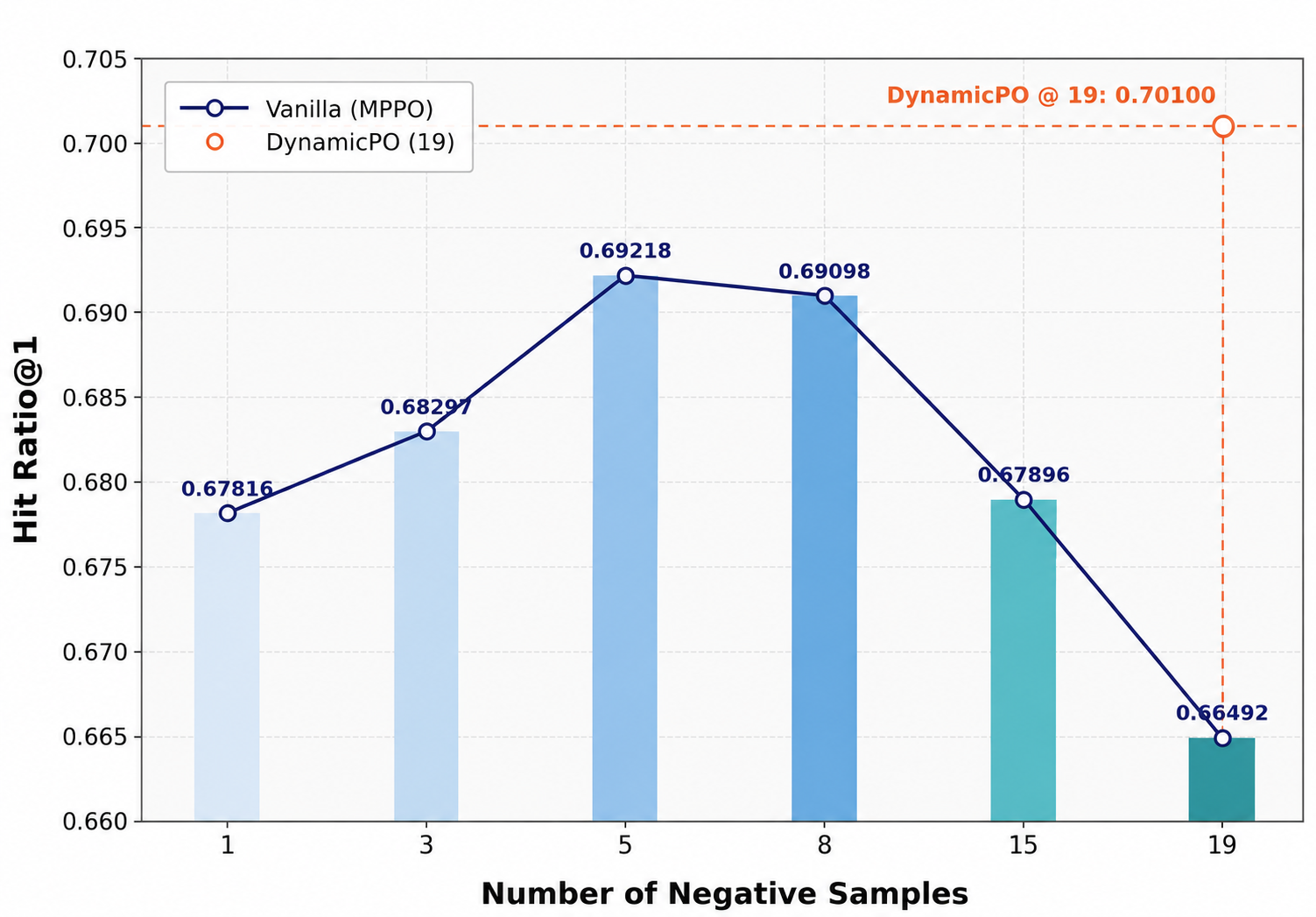}
\caption{Scaling behavior of MPPO as the number of negative samples increases.}
\label{fig:mppo_negative_scaling}
\end{figure}

\paragraph{\textbf{Preference optimization collapse extends across multi-negative DPO formulations.}}
Figures~\ref{fig:dmpo_negative_scaling} and~\ref{fig:mppo_negative_scaling} reveal a shared non-monotonic scaling pattern. DMPO improves from 0.6218 with one negative to 0.6332 with three negatives, \textbf{but deteriorates rapidly as the negative set continues to grow,} decreasing to 0.5847 with 15 negatives. MPPO tolerates a larger negative set, reaching its best performance of 0.6922 with five negatives and remaining close to this peak with eight negatives, \textbf{but eventually declining to 0.6649 with 19 negatives.} These results demonstrate that \textbf{preference optimization collapse is not specific to DMPO, although its onset and severity vary across objective formulations}: DMPO exhibits an earlier and sharper collapse, whereas MPPO delays and moderates, but does not eliminate, the degradation induced by excessive negatives. \textbf{This objective-dependent behavior is consistent with our gradient-suppression analysis}: as the negative set expands, the accumulated contribution of model-discriminative negatives can dilute the relatively sparse supervision provided by boundary-critical negatives, thereby weakening preference-boundary refinement. DynamicPO directly alleviates this issue by dynamically identifying boundary-critical negatives and adaptively calibrating their optimization strength. Under the 19-negative setting, it improves DMPO from 0.5936 to 0.6854 and MPPO from 0.6649 to 0.7010, \textbf{both exceeding the best performance achieved by their corresponding vanilla scaling curves.} This result indicates that the observed degradation is not caused by the enlarged negative pool itself, but by the indiscriminate treatment of negatives with substantially different boundary informativeness.

\begin{figure}[H]
\centering
\includegraphics[width=0.68\linewidth]{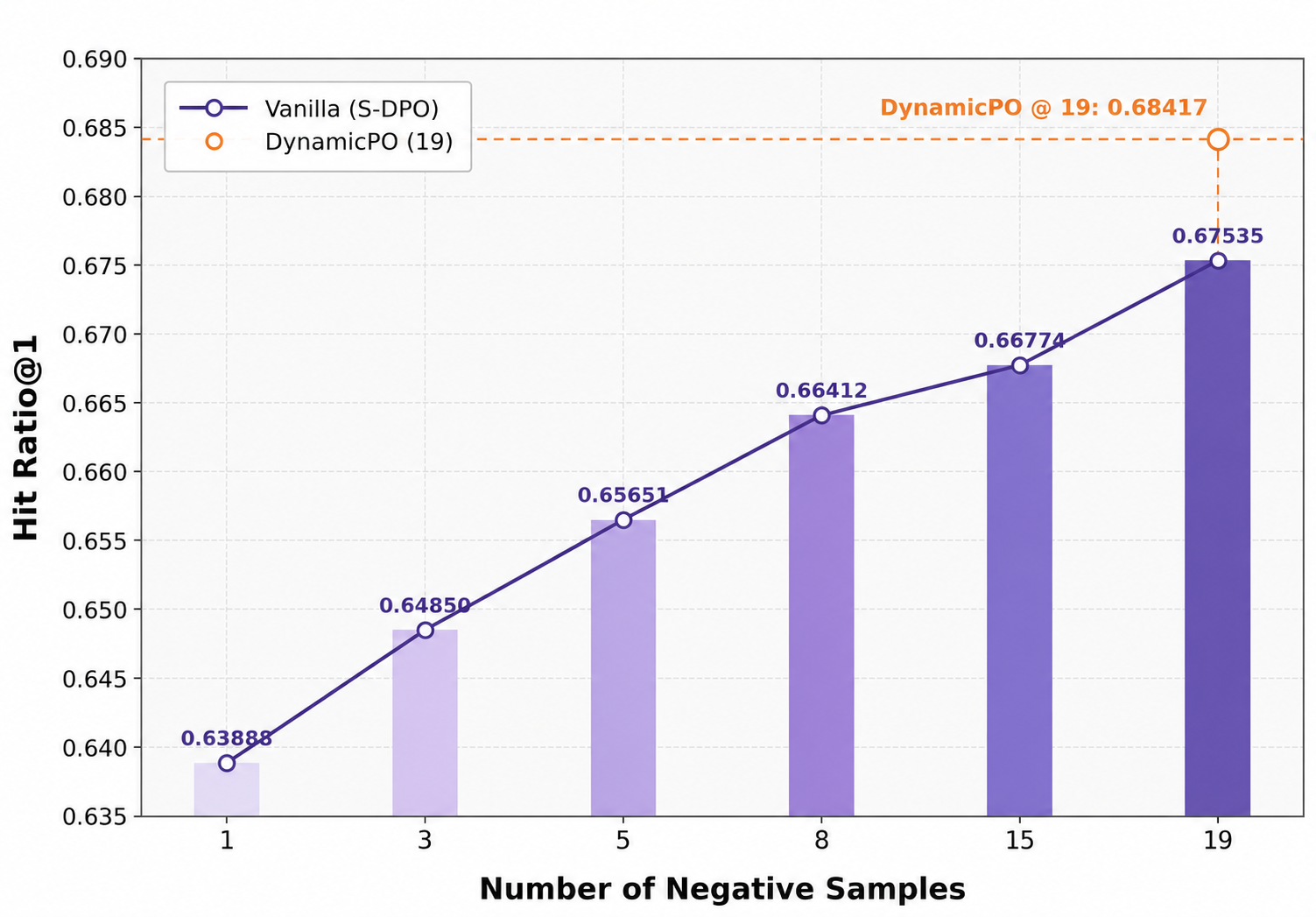}
\caption{Scaling behavior of S-DPO as the number of negative samples increases.}
\label{fig:sdpo_negative_scaling}
\end{figure}

\paragraph{\textbf{S-DPO remains stable, but stability does not imply objective superiority.}}
As shown in Figure~\ref{fig:sdpo_negative_scaling}, S-DPO exhibits a more stable scaling pattern than DMPO and MPPO, with its performance increasing consistently from 0.6389 with one negative to 0.6754 with 19 negatives and no visible collapse within the evaluated range. 
S-DPO may implicitly attenuate the contribution of negatives that are already well separated from the positive item, thereby reducing its sensitivity to the accumulation of model-discriminative negatives. 
\textbf{However, this scaling stability does not imply that S-DPO fully exploits informative negative supervision, nor does it establish its uniform superiority over other multi-negative DPO objectives.}
Integrating DynamicPO with S-DPO further improves its performance from 0.6754 to 0.6842. Moreover, applying DynamicPO to DMPO and MPPO yields 0.6854 and 0.7010, respectively.
Consequently, applying DynamicPO to all three objectives yields performance above vanilla S-DPO under the same 19-negative setting.
These results indicate that \textbf{scaling robustness and optimization effectiveness are related but distinct properties. }
S-DPO provides stronger inherent robustness to enlarged negative sets, whereas DynamicPO enables DMPO and MPPO to realize greater optimization potential by explicitly identifying and strengthening boundary-critical supervision.
By dynamically identifying boundary-critical negatives and adaptively calibrating their optimization strength, DynamicPO complements these objective-specific properties and preserves informative boundary supervision across all three objectives.

\subsection{Future Directions}
\label{app:future_directions}

Our results demonstrate that dynamically selecting boundary-critical negatives and adaptively calibrating their optimization strength can improve multi-negative preference optimization for LLM-based recommendation. These findings also motivate several directions for further investigation.

\paragraph{\textbf{Characterizing dynamic-$\beta$ mechanisms.}}
The dynamic-$\beta$ mechanism in DynamicPO performs sample-specific calibration of optimization strength based on the likelihood gap between the preferred sample and each selected boundary-critical negative. Although this mechanism consistently improves different multi-negative objectives, its optimization behavior is not yet fully understood. Future work may characterize how dynamic $\beta$ affects gradient allocation, training stability, and preference-boundary refinement, and examine how these effects vary across objective formulations, negative-set sizes, and model scales. Such analysis could provide a more principled basis for designing adaptive optimization-strength rules under different recommendation settings.

\paragraph{\textbf{Extending DynamicPO beyond recommendation.}}
This work focuses on recommendation, where preference ambiguity can be characterized by the likelihood gaps between preferred and negative items. The same principle may extend to broader LLM alignment tasks, such as dialogue, question answering, and instruction following, where rejected responses can also differ substantially in boundary informativeness. Extending DynamicPO to these settings requires task-specific definitions of boundary-critical responses and corresponding mechanisms for calibrating their optimization strength. Although $\beta$-DPO~\cite{betadpo} investigates dynamic $\beta$ adjustment for natural-language generation, directly applying its calibration strategy to recommendation leads to degraded performance in our experiments. This result indicates that adaptive-$\beta$ mechanisms may not transfer unchanged across preference optimization scenarios: the appropriate calibration granularity, direction, and magnitude can depend on the target task. In contrast, DynamicPO jointly identifies boundary-critical negatives and calibrates their optimization strength at the sample level according to boundary ambiguity. Future work may build on these observations to investigate whether and how boundary-aware sample selection and adaptive optimization-strength calibration can benefit broader LLM alignment tasks.

\paragraph{\textbf{Extending DynamicPO to online RL.}}
DynamicPO is developed for multi-negative DPO objectives, which extend DPO, an offline preference optimization method derived from a KL-regularized reward-maximization objective.
It learns from pre-collected preference comparisons without fitting an explicit reward model or collecting new rollouts from the current policy. Our findings show that even in this offline setting, indiscriminately optimizing all negatives can dilute boundary-critical supervision and induce preference optimization collapse. A natural next step is therefore to examine whether the same boundary-aware principle can benefit online RL methods such as PPO and GRPO, where training samples are continuously generated by the current policy and can differ substantially in their utility for policy improvement.

In RL-based generative recommendation, sampled outputs or ranked lists may differ not only in reward or advantage, but also in how effectively they resolve uncertain user preferences. Some rollouts may be invalid, redundant, or already clearly inferior, whereas others may receive similar rewards or remain close to the current ranking boundary and thus provide more informative supervision. Extending DynamicPO to this setting would require redefining boundary-critical samples in terms of online reward, advantage, and ranking uncertainty. Boundary-aware rollout selection could then prioritize samples that expose unresolved preference distinctions, while sample-specific update-strength calibration could control their contribution to policy optimization. Investigating whether these mechanisms can prevent informative recommendation rollouts from being overwhelmed by less useful samples is a promising direction for future work.

\subsection{Reproduction Results on NVIDIA H200 GPUs} \label{app:h200_reproduction}

Additionally, we reproduce the main LastFM experiments on NVIDIA H200 GPUs. We evaluate DynamicPO across three multi-negative preference optimization objectives with Llama2-7B, and across three backbone language models with DMPO as the base objective. 
As shown in Tables~\ref{tab:h200_objective_reproduction} and~\ref{tab:h200_backbone_reproduction}, DynamicPO consistently outperforms the corresponding vanilla objectives across all objectives and backbone models, confirming that its effectiveness is not specific to a particular objective formulation or language model. The reproduced checkpoints are publicly released on the  \href{https://huggingface.co/xingyuHuxingyu/DynamicPO}{Hugging Face Model Hub}.

\begin{table}[H]
\centering
\caption{HitRatio@1 of vanilla and DynamicPO variants across multi-negative objectives on LastFM.}
\label{tab:h200_objective_reproduction}
\small
\begin{adjustbox}{max width=0.65\textwidth}
\begin{tabular}{lcc}
\toprule
\textbf{Objective} & \textbf{Vanilla} & \textbf{DynamicPO} \tabularnewline
\midrule
DMPO  & 0.58757 & \cellcolor{gray!8}\textbf{0.67535} \tabularnewline
MPPO  & 0.67454 & \cellcolor{gray!8}\textbf{0.69419} \tabularnewline
S-DPO & 0.66774 & \cellcolor{gray!8}\textbf{0.67575} \tabularnewline
\bottomrule
\end{tabular}
\end{adjustbox}
\end{table}

\begin{table}[H]
\centering
\caption{HitRatio@1 of vanilla DMPO and DynamicPO across backbone language models on LastFM.}
\label{tab:h200_backbone_reproduction}
\small
\begin{adjustbox}{max width=0.75\textwidth}
\begin{tabular}{lcc}
\toprule
\textbf{Backbone} & \textbf{Vanilla} & \textbf{DynamicPO} \tabularnewline
\midrule
Llama2-7b-hf        & 0.58757 & \cellcolor{gray!8}\textbf{0.67535} \tabularnewline
Llama3-8B-Instruct  & 0.60481 & \cellcolor{gray!8}\textbf{0.73106} \tabularnewline
Qwen2.5-7B-Instruct & 0.56874 & \cellcolor{gray!8}\textbf{0.64529} \tabularnewline
\bottomrule
\end{tabular}
\end{adjustbox}
\end{table}

\bibliographystyle{splncs04}
\bibliography{references}

\end{document}